\documentstyle[prb,multicol,epsf,aps]{revtex}

\begin{document}

\title{Coupling of Surface Acoustic Waves to a Two Dimensional
  Electron Gas}

\author{Steven H.  Simon} \address{Department of Physics,
  Massachusetts Institute of Technology, Cambridge, MA 02139.}

%\date{\today}
\maketitle

\begin{abstract}
  When a surface acoustic wave (SAW) is coupled piezoelectrically to a
  two dimensional electron gas (2DEG), a velocity shift and
  attenuation of the SAW are induced that reflect the conductivity of
  the 2DEG.  This paper considers the case of a AlGaAs heterostructure
  with a 2DEG a distance $d$ from a (100) surface of the crystal where
  the SAWs are propagated in the [011] direction at wavevector $q$.
  It is found that the velocity shift $\Delta v_s$ and the attenuation
  coefficient $\kappa$ satisfy the well known equation $(\Delta
  v_s/v_s) - (i \kappa/q) = (\alpha^2/2)/\left(1 + \frac{i
    \sigma_{xx}(q,\omega)}{\sigma_m}\right)$ where
  $\sigma_{xx}(q,\omega)$ is the complex conductivity at wavevector
  $q$ and frequency $\omega = v_s q$ with $v_s$ the velocity of the
  SAW.  The coefficients $\alpha$ and $\sigma_m$ are calculated and it
  is found that $\alpha$ has a nontrivial dependence on the product
  $qd$.
\end{abstract}

\vspace*{-3in} \hspace*{5in} \begin{minipage}{4in} \small Submitted to
Phys Rev B \\ May 6, 1996 \end{minipage} \vspace*{2.5in}

\begin{multicols}{2}

\section{Introduction}

For almost thirty years it has been known that the velocity $v_s$ of
surface acoustic waves (SAWs) in piezoelectric crystals can be
effected by the electrical properties of nearby
conductors\cite{Inge1,Tseng}.  If the nearby conductors are
dissipative, then they can allow the SAW to attenuate also.  Work by
Ingebrigtsen\cite{Inge1}, and later authors\cite{Bier,Efros} showed
that when a piezoelectric is brought next to a thin layer of a
conducting medium, the SAW velocity shift $\Delta v_s$ and the
attenuation coefficient $\kappa$ satisfy the relation
\begin{equation}
  \label{eq:mastersaw}
  \frac{\Delta v_s}{v_s} - \frac{i \kappa}{q} = \frac{\alpha^2/2}{1 +
    i \sigma_{xx}(q,\omega)/\sigma_m}
\end{equation}
where $\sigma_{xx}(q,\omega)$ is the longitudinal conductivity of the
adjoining medium at wavevector $q$ and frequency $\omega = v_s q$.
Note that the velocity shift is measured with respect to the velocity
of the SAW when the adjoining medium has infinite conductivity.  The
coefficients $\sigma_m$ and $\alpha^2/2$ depend on material parameters
and are discussed at length in this paper.

Using the above relation between SAW velocity shift (or attenuation)
and the conductivity of a surface layer, experimentalists have probed
the conducting properties of two dimensional electron gases (2DEGs)
placed near the surface of crystals of
GaAs\cite{Weimann,Wixforth,Willett1,Willett2,Willett3,NewWillett,Others}.
(An approximation of the experimental geometry is shown in Fig. 1.)
In the earlier of these experiments\cite{Weimann,Wixforth}, the
wavelength of the probing SAW was much larger than the distance $d$
from the surface.  In this case, the depth $d$ can be neglected, and
the coefficients $\sigma_m$ and $\alpha^2/2$ can be assumed to be
constant.  However, in the more recent
experiments\cite{Willett2,Willett3,NewWillett,Willettcom}, the
wavevector $q$ can be so large that the product $qd$ is or order
unity.  In this case, one must carefully consider the wavevector
dependencies of these coefficients.  Roughly one might expect that the
coupling $\alpha^2/2$ should decay approximately as $e^{-2 q d}$.
However, it is seen experimentally that the coupling remains roughly
constant up to the highest wavevectors probed ($qd \approx 4$).  In
this paper, the wavevector dependences of $\sigma_m$ and $\alpha^2/2$
are explicitly derived for an experimental geometry similar to that
used in these experimental works.  Using the results derived here, it
should be possible to deduce quantitative results about the frequency
and wavevector dependent conductivities of the samples (A detailed
analysis of the data in References \onlinecite{Willett3} and
\onlinecite{NewWillett} is given in Reference \onlinecite{Simonwave}).

In References
\onlinecite{Weimann,Wixforth,Willett1,Willett2,Willett3,NewWillett,Willettcom},
the SAWs are propagated in the [011] direction along a (100) surface
of an (Al)GaAs crystal.  For most of this paper it will be sufficient
to approximate this system as the geometry shown in Fig. 1.  In other
words we assume that the 2DEG is a thin conducting layer a distance
$d$ (typically between 1000 and 5000 Angstroms) from the surface of a
homogeneous ${\mbox{Al}_x\mbox{Ga}_{1-x}\mbox{As}}$ crystal with the
fraction $x$ of Al taken to be approximately $30\%$.  The effects of
the differences between the actual experiments and this idealization
will be considered in the concluding section of this paper.

The outline of the remainder of this paper is as follows.  In section
\ref{sec:response}, the electromagnetic response function $K_{00}$ is
defined and related to the conductivity $\sigma_{xx}$.  The parameter
$\sigma_m$ is then defined in terms of the SAW velocity and the
effective dielectric constant (ie the effective strength of the
Coulomb interaction in the 2DEG).  The effective dielectric constant
is calculated explicitly in the appendix.  In section \ref{sec:diss} a
qualitative explanation is given as to how the SAW induces a potential
through piezoelectric coupling, thus resulting in an energy shift
and/or dissipation through the real and/or imaginary part of the
conductivity.  Equation \ref{eq:mastersaw} is then derived leaving
only the coupling constant $\alpha^2/2$ to be calculated.  In section
\ref{sec:details} the form of the SAW (neglecting the piezoelectric
coupling) is discussed with particular focus on finding the energy
density per unit area of the SAW.  The effect of a small piezoelectric
coupling is then considered in section \ref{sec:piezo} yielding a form
for the induced potential.  The coupling $\alpha^2/2$ is then derived
and is found to have a nontrivial and nonmonotonic dependence on $qd$.
Finally, section \ref{sec:conclusion} considers a number of
experimental issues and summarizes results.

\end{multicols}
\begin{figure}[htbp]
  \begin{center}
    \leavevmode
    \epsfxsize=5in
    \epsfbox{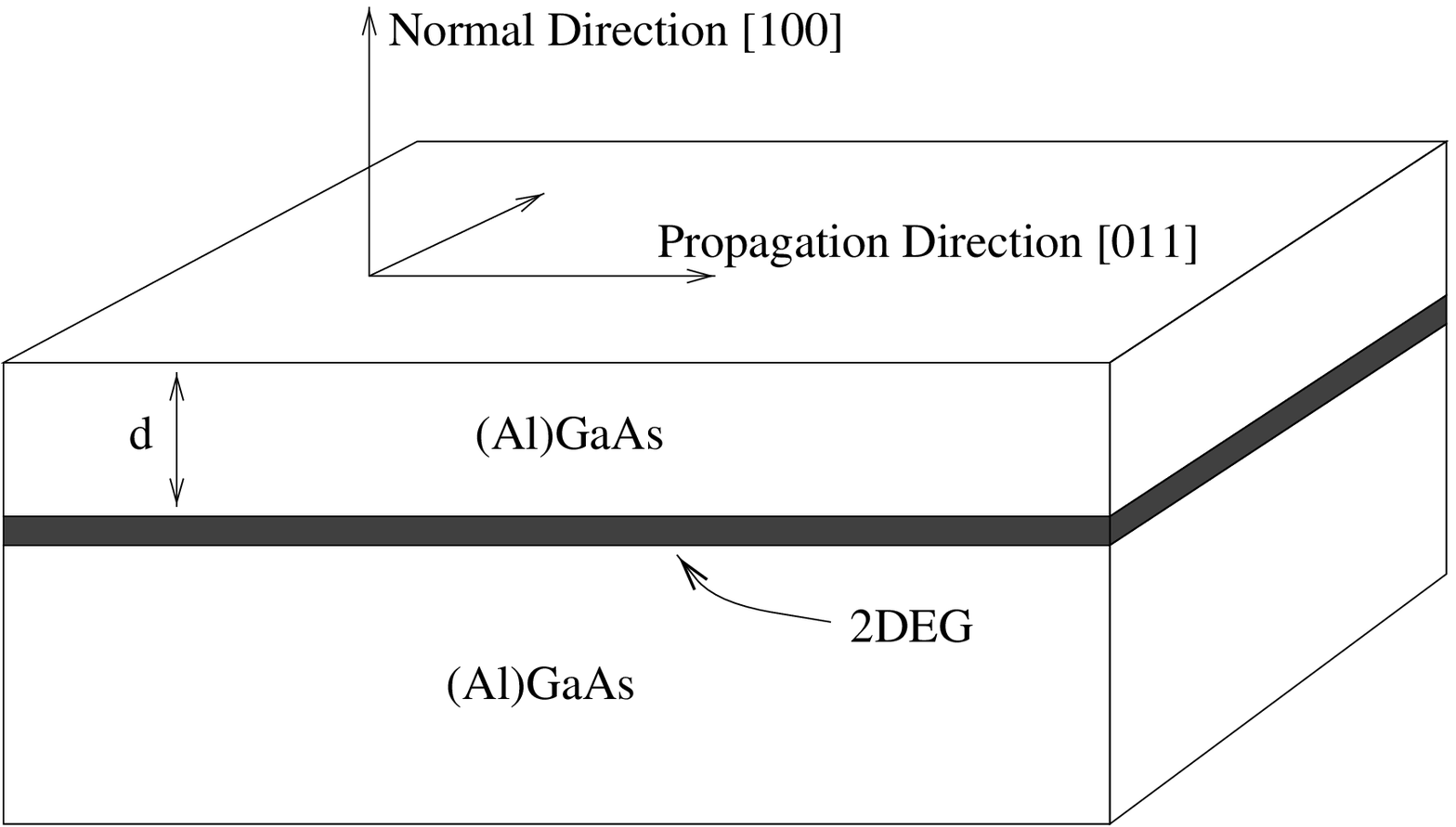}
  \end{center}
  {\centerline{\begin{minipage}[t]{5in} \small{Fig. 1: Model Geometry
        for Surface Acoustic Wave Experiments.  In the experiments the
        spacing $d$ is typically between 1000 and 5000 Angstroms.  The
        ${\mbox{Al}_x\mbox{Ga}_{1-x}\mbox{As}}$ typically has a
        fraction of Al given by $x\approx 30 \%.$}
      \end{minipage}}}
  \label{fig:geom}
\end{figure}
\begin{multicols}{2}

\section{Response Functions}
\label{sec:response}

The density-density response $K_{00}(q,\omega)$ is defined by the
relation
\begin{equation}
  \label{eq:Kdef}
  n(q,\omega) = K_{00}(q,\omega)
  \phi^{\mbox{\scriptsize{ext}}}(q,\omega)
\end{equation}
where $\phi^{\mbox{\scriptsize{ext}}}$ is the perturbing externally
applied scalar potential applied at a frequency $\omega$ and
wavevector ${\bf q} = q {\hat{\bf x}}$, and $n(q,\omega)$ is the
induced fluctuation density.  As we will see below, the SAW
experiments directly measure $K_{00}$ at finite frequency and
wavevector.

Many linear response measurements do not measure, however, the ratio
of induced density to the externally applied potential but rather the
response to the total potential.  A density $n({\bf q})$ induced by
the external vector potential, gives rise to a Coulomb scalar
potential
\begin{equation}
  \label{eq:vq}
  \phi^{\mbox{\scriptsize{ind}}}(q,\omega) = - v(q) n(q,\omega)
\label{coulomb}
\end{equation}
where $v(q) = \frac{2\pi}{{\epsilon_{\mbox{\tiny{eff}}}} q}$ is the
Fourier transform of the usual Coulomb interaction $ v(r) =
1/({\epsilon_{\mbox{\tiny{eff}}}} |r|)$.  (In principle, currents in
the sample give rise to an induced vector as well as scalar potential,
but in practice these fields are negligible).  Here
${\epsilon_{\mbox{\tiny{eff}}}}$ is the effective background
dielectric constant.  The wavevector dependent form of this dielectric
constant for the case of the model geometry of Fig. 1 is derived in
the appendix, and is given by
\begin{equation}
  \label{eq:eeff}
  \frac{{\epsilon_{\mbox{\tiny{eff}}}}}{\epsilon} = \frac{1}{2} \left(
  \frac{(\epsilon+\epsilon_0) \exp(qd)}{ \epsilon \cosh(qd) +
    \epsilon_0 \sinh(qd)} \right)
\end{equation}
where $\epsilon$ is the dielectric constant of the bulk
${\mbox{Al}_x\mbox{Ga}_{1-x}\mbox{As}}$ and $\epsilon_0$ is the
dielectric constant of the medium above the surface ($\approx 1$).
The dielectric constant\cite{Optical} for
${\mbox{Al}_x\mbox{Ga}_{1-x}\mbox{As}}$ with $x\approx .3$ is
approximately 12.5 ( which is slightly lower than the dielectric
constant for GaAs which is approximately 13.0).

Using Eq. \ref{eq:vq}, the total scalar potential
\begin{equation}
  \phi^{\mbox{\scriptsize{tot}}} = \phi^{\mbox{\scriptsize{ext}}} +
  \phi^{\mbox{\scriptsize{ind}}}
\end{equation}
is now written as
\begin{equation}
  \label{eq:exttot}
  \phi^{\mbox{\scriptsize{tot}}} = ( 1 - v(q) K_{00})
  \phi^{\mbox{\scriptsize{ext}}}. 
\end{equation}
It then becomes useful to define the polarization
$\Pi_{00}(q,\omega)$, which relates the induced density $n(q,\omega)$
to the total scalar potential via
\begin{equation} 
  n(q, \omega) = \Pi_{00}(q, \omega)
  \phi^{\mbox{\scriptsize{tot}}}(q,\omega)
\end{equation}
Combining this definition with Eqs. \ref{eq:Kdef} and \ref{eq:exttot}
yields the  equation
\begin{equation}
  \label{eq:K00Pi00}
  [K_{00}]^{-1} = \left[\Pi_{00}\right]^{-1} + v(q).
\end{equation}
Since the response function $\Pi_{00}$ relates the density to the
total vector potential, it is useful to write this function in terms
of the conductivity $\sigma_{\alpha\beta}$ which relates the two
spatial components of the current $(j_x,j_y)$ to the two spatial
components of the total electric field $(E_x,E_y)$ via ${\bf j} =
\sigma {\bf E}$.  Using current conservation to give $j_x = (\omega
n/q)$, then yields
\begin{equation}
  \sigma_{xx} = \frac{-i \omega}{q^2} \Pi_{00}
\end{equation}
In particular, this allows us to write the general relation
\begin{equation}
  \label{eq:K00sigmaxx}
  [K_{00}]^{-1} =  v(q) - \frac{i \omega}{q^2 \sigma_{xx}}
\end{equation}
Throughout this work, we will assume that $\omega = v_s q$ with $v_s$
the SAW velocity, which then implies
\begin{equation}
  K_{00}(q,\omega) = \frac{{\epsilon_{\mbox{\tiny{eff}}}} q}{2 \pi(1 -
    i \sigma_{m}/\sigma_{xx}(q,\omega))}
\end{equation}
with 
\begin{equation}
  \sigma_m = \frac{v_s {\epsilon_{\mbox{\tiny{eff}}}}}{2 \pi}.
\end{equation}
The function $\sigma_m(qd)$ is shown in Fig. 2.  Here, the
experimentally relevant parameters for References
\onlinecite{Willett3} and \onlinecite{NewWillett} are used.  These are
$\epsilon = 12.5$ and $v_s = 3010 m/sec$ (see section
\ref{sec:details} below).

\section{Induced Energy Shift}
\label{sec:diss}

Due to the piezoelectric coupling, an external scalar potential
$\phi^{\mbox{\scriptsize{ext}}}$ is induced in the 2DEG.  For now, we
will write
\begin{equation} 
  \label{eq:aex}
  \phi^{\mbox{\scriptsize{ext}}} =  C e_{14} F(qd)/\epsilon
\end{equation}
where $C$ is the amplitude of the SAW, $e_{14}$ the piezoelectric
stress constant, and $F$ is a dimensionless function of $qd$ that
represents the fact that the SAW decays into the bulk.  Clearly $F$
should approach a constant as $qd \rightarrow 0$ and should approach
zero as $qd \rightarrow \infty$.  Roughly, one should expect that the
function $F$ should decay as $e^{-qd}$.

The induced energy density per unit area due to this external
potential is given by
\begin{equation}
  \delta U = \frac{1}{2} K_{00} |\phi^{\mbox{\scriptsize{ext}}}|^2
\end{equation}
This expression is obtained from integrating a differential $d \delta
U = n(\phi^{\mbox{\scriptsize{ext}}}) d
\phi^{\mbox{\scriptsize{ext}}}$ and using Eq. \ref{eq:Kdef}. (Note
that using $\phi^{\mbox{\scriptsize{tot}}}$ here instead would account
for only the electrical energy).  Using Eq. \ref{eq:K00sigmaxx} we can
rewrite this shift as
\begin{equation}
  \delta U = \frac{{\epsilon_{\mbox{\tiny{eff}}}} q}{4 \pi (1 - i
    \sigma_{m}/\sigma_{xx}(q,\omega))}
  |\phi^{\mbox{\scriptsize{ext}}}|^2
\end{equation}
We now want to measure this energy shift with respect to the shift for
$\sigma_{xx} \rightarrow \infty$.  Thus,
\begin{eqnarray}
  \Delta U &\equiv& \delta U - \delta U(\sigma_{xx} = \infty) \\ &=&
  \frac{{\epsilon_{\mbox{\tiny{eff}}}} q}{4 \pi } \left[ \frac{1}{1 -
    i \sigma_{m}/\sigma_{xx}(q,\omega)} - 1 \right]
  |\phi^{\mbox{\scriptsize{ext}}}|^2 \\ & = &
  \frac{{\epsilon_{\mbox{\tiny{eff}}}} q}{4 \pi } \left[ \frac{1}{1 +
    i \sigma_{xx}(q,\omega)/\sigma_m } \right]
  |\phi^{\mbox{\scriptsize{ext}}}|^2
\end{eqnarray}

It is found below that the surface acoustic wave has an energy density
proportional to $C^2 q^2$ where $C$ is the amplitude of the wave and
$q$ is the wavevector.  Furthermore, the wave decays exponentially
into the bulk with a decay constant proportional to $q$.  Thus, when
integrated in the ${\hat{\bf z}}$ direction, the energy $U$ per unit
surface area is given by
\begin{equation}
  \label{eq:energyden}
  U = q C^2 H
\end{equation}
where $H$ is a factor that depends on material parameters that we will
determine below.  Combining this with the results of the above
section, the fractional energy shift is then given by
\begin{equation}
  \frac{\Delta U}{U} = 
  \frac{\alpha^2/2}{1 + i \sigma_{xx}(q,\omega)/\sigma_m } 
\end{equation}
where 
\begin{equation}
  \label{eq:msalph}
  \frac{\alpha^2}{2} = \frac{{\epsilon_{\mbox{\tiny{eff}}}}}{\epsilon}
  \frac{e_{14}^2 }{4 \pi \epsilon H} |F(qd)|^2
\end{equation}
(Note that the factor of $4 \pi$ will vanish when
${\epsilon_{\mbox{\tiny{eff}}}}$ is converted into MKSI units).
Clearly, this result implies the velocity shift and attenuation
relation given by Eq.  \ref{eq:mastersaw}.  All that now remains is the
tedious job of evaluating the constant $H$ as well as the functional
form $F$.  It should be noted that in the small $qd$ limit, various
experiments\cite{Grudowski,Weimann,Wixforth,Willett1,Willett2,Willett3}
have measured the value of the coupling constant and have found
$\alpha^2/2 \approx 3.2 \times 10^{-4}$.  As is discussed below in
section \ref{sec:conclusion}, these measurements should be viewed with
caution.  As discussed above, one expects roughly that $F$ decays as
$e^{-qd}$ so that $\alpha^2/2$ decays as $e^{-2 qd}$.  This, however,
contradicts experimental observation\cite{Willettcom}.  Below, in a
more careful analysis, we will see why the decay is actually somewhat
slower and shows a nonmonotonic dependence on $qd$.

\section{Non-Piezoelectric SAWs}
\label{sec:details}

We begin by discussing the solution of the SAW equations with the
piezoelectric coupling set to zero.  The piezoelectric coupling will
then be added at lowest order.

Defining a displacement vector $u_k$, the elastic wave equation is
given by\cite{Landau,Farnell,Auld}
\begin{equation}
  \label{eq:waveeq}
  c_{ijkl} \partial_l \partial_i u_k + \rho \ddot u_j = 0
\end{equation}
where $\rho$ is the mass density, $c$ is the elastic tensor, we have
used the notations $\partial_l f= \frac{\partial f}{\partial x_l}$, ,
$\dot{f} = \frac{\partial f}{\partial t}$, and repeated indices are
summed. For GaAs, AlAs, and other crystals of cubic symmetry there are
only 3 independent elastic constants.  These constants are
conventionally called $c_{11}$, $c_{12}$, and $c_{44}$.  For GaAs at
low temperatures, the elastic constants $c_{11}$, $c_{12}$, and
$c_{44}$ are given by\cite{Handbook} $12.26 \times 10^{10}, 5.71
\times 10^{10}$, and $6.00 \times 10^{10} N/m^2$ respectively.  The
constants for AlAs\cite{Handbook,Chetty} are given approximately by
$12.2 \times 10^{10}, 5.5 \times 10^{10}$, and $5.7 \times 10^{10}
N/m^2$ respectively.  It is noted that the elastic constants of the
two materials are roughly the same.  For
${\mbox{Al}_x\mbox{Ga}_{1-x}\mbox{As}}$ it is reasonable to
interpolate for any value of $x$.  (Experimentally, there may be some
uncertainty in $x$).  The density\cite{Handbook} of GaAs is 5307
$kg/m^3$, and the density of AlAs is 3598 $kg/m^3$.  Thus for
${\mbox{Al}_x\mbox{Ga}_{1-x}\mbox{As}}$ with $x \approx .3$, the
density interpolates to approximately 4794 $kg/m^3$, which differs
from that of GaAs by only 10\%.

In considering surface waves, the wave equation must be supplemented
with the boundary condition at the free surface that there is no total
force at the surface.  This condition is written
as\cite{Landau,Farnell}
\begin{equation}
\label{eq:boundary}
c_{{\hat{\bf z}}jkl} \partial_l u_k = 0.
\end{equation}
where the subscript ${\hat{\bf z}}$ represents the direction normal to
the surface.  For certain geometries, analytic solutions of the SAW
equations are available.  In the present case of a (100) surface with
wave propagation in the [011] direction, the velocity of SAW
propagation is given by the solution (here we are interested in the
lowest velocity solution) of the cubic equation\cite{Stonely}.
\begin{equation}
  \left( 1 - \frac{c_{11}}{c_{44}} X \right)\left( \frac{c_{11}
      c_{11}' - c_{12}^2}{c_{11}^2} - X \right)^2 = X^2 \left(
    \frac{c_{11}'}{c_{11}} - X \right)
\end{equation}
where $c_{11}' = \frac{1}{2} ( c_{11} + c_{12} + 2 c_{44})$ and $X =
\rho v_s^2/c_{11}$ gives the SAW velocity $v_s$.

For ${\mbox{Al}_x\mbox{Ga}_{1-x}\mbox{As}}$ with $x \approx .3$, the
velocity is approximately 3010 m/sec.  (This differs from that of pure
GaAs by only 5\%.)  Once the velocity is determined, one can easily
solve analytically for the form of the SAW.  For the experimental
geometry we are presently considering, the displacements for this wave
can be written as\cite{Stonely,endnote2}
\begin{eqnarray}
  \label{eq:uform}
  u_x &=& C \left(e^{-\Omega q z - i  \varphi} + e^{-\Omega^* q z + i 
    \varphi} \right) e^{iq (x - v_s t)}
\\i  u_z &=& C \left(\gamma e^{-\Omega q z - i  \varphi} + \gamma^*
e^{-\Omega^* q z + i  \varphi} \right) e^{iq (x- v_s t)}
\end{eqnarray}
with $u_y = 0$ and $C$ the amplitude ($C$ has dimensions of length).
Here, the ${\hat{\bf x}}$ direction is chosen in the direction of wave
propagation (the [011] direction).  The parameters $\Omega, \gamma$,
and $\varphi$ are determined by\cite{Stonely,endnote2}
\begin{eqnarray} \nonumber
  0 &=& (c_{11}' - X c_{11} - \Omega^2 c_{11})(c_{44} - X c_{11} -
  \Omega^2 c_{44})\\ & & ~~~~~~~~~~~~~~ + \Omega^2 (c_{12} + c_{44})^2
 \\ \gamma &=& \Omega  \label{eq:Omegaroot}
 \left[ \frac{c_{12} + c_{44}}{c_{44} - (X + \Omega^2) c_{11}}\right]
    \\ e^{-2 i \varphi} &=& - \frac{\gamma^* - \Omega^*}{\gamma - \Omega}
\end{eqnarray}
In the case of ${\mbox{Al}_x\mbox{Ga}_{1-x}\mbox{As}}$ with $x = .3$, the values of $\gamma$ and
$\Omega$ and $\varphi$ are given in this case by $\Omega \approx .501
+ .472 i$, $\gamma \approx -.705 + 1.146i$, and $\phi \approx 1.06$.

The local energy density of this wave can be written
as\cite{Landau,Auld}
\begin{equation}
  E = \frac{1}{2} c_{ijkl} u_{ij} u_{kl}^*
\end{equation}
where the strain $u_{ij}$ is given by
\begin{equation}
  u_{ij} = \frac{1}{2} (\partial_i u_j + \partial_j u_i)
\end{equation}
For the AlGaAs surface wave discussed above, the energy density can be
written as\cite{Auld}
\begin{eqnarray}
  \label{eq:eden2} \nonumber
  E &=& \frac{1}{2} \left( c_{11}' |u_{xx}|^2 + c_{11} |u_{zz}|^2
\right. \\  & & ~~~~~~~~~~ \left. +2 c_{12} Re[u_{xx}^* u_{zz}] + c_{44} |2
u_{xz}|^2 \right)
\end{eqnarray}
Inserting the above described form of the wave, yields the strains
\begin{eqnarray}
  u_{xx} &=&  i q u_x
  \\
  u_{zz} &=& i q C \left( \gamma \Omega e^{-\Omega q z - i \varphi} +
  \mbox{c.c.} \right)e^{iq(x - v_s t)} \\ 2  u_{xz} &=& C q \left(
  [\gamma - \Omega] e^{-\Omega q z - i \varphi} + \mbox{c.c.}
\right)e^{iq(x - v_s t)}
\end{eqnarray}
with ``c.c.'' meaning complex conjugate.  Finally, integrating the result
of Eq.  \ref{eq:eden2} in the ${\hat{\bf z}}$ direction yields an energy per
unit surface area in the form given by Eq.  \ref{eq:energyden} with
\begin{eqnarray}
  \nonumber H &=& \mbox{Re} \left[ c_{11}' \left( \frac{e^{-2 i
      \varphi}}{\Omega} + \frac{1}{\alpha} \right) \right. \\ 
  \nonumber &+& c_{11} \left( \frac{(\gamma - \Omega)^2 e^{-2 i
      \varphi}}{\Omega} + \frac{|\gamma - \Omega|^2}{\alpha} \right )
  \\ \nonumber &+& c_{44} \left( (\gamma^2 \Omega) e^{- 2 i \varphi}+
  \frac{|\gamma \Omega|^2}{\alpha} \right ) \\ &+& \left. 2 c_{12}
  \left( \gamma e^{-2 i \varphi} + \frac{\mbox{Re}(\gamma
      \Omega)}{\alpha} \right) \right]
\end{eqnarray}
where $\alpha = \mbox{Re}(\Omega)$.  This yields a numerical value of 
\begin{equation}
H \approx 28.8 \times 10^{10} N/m^2
\end{equation}
(The value for pure GaAs is lower by only about 2 \%).

\section{Piezoelectric Coupling}
\label{sec:piezo}

When a piezoelectric coupling is added, the wave equations take the
form\cite{Inge1,Farnell,Auld}
\begin{eqnarray}
  \label{eq:waveeq2}
  c_{ijkl} \partial_l \partial_i u_k + e_{kij} \partial_k \partial_i
  \phi + \rho \ddot u_j &=& 0 \\
  e_{ikl} \partial_l \partial_i u_k - \epsilon \nabla^2 \phi = 0
  \label{eq:Poisson} 
\end{eqnarray}
where $e$ is the piezoelectric stress tensor, $\phi$ is the electric
potential, and $\epsilon$ is the dielectric constant of the medium
(here $\epsilon$ is assumed to be isotropic).  For GaAs, AlAs (and
other cubic crystals of class $\bar{4}3m$), there is only one
independent nonzero component of the piezoelectric tensor\cite{Auld}
called $e_{14}$.

The value of $e_{14}$ for GaAs has an accepted
value\cite{Handbook,Auld} of approximately .157 $C/m^2$.  However, it
should be noted that there is a small amount of evidence\cite{Smaller}
that the actual value might be somewhat lower (by perhaps as much as
40\%).  For the present work we will choose to work with the accepted
value.  For the case of AlAs, it is even more difficult to find a
reliable value for the piezoelectric coupling.  To the author's
knowledge, no reliable measurement of this quantity has been made to
date\cite{IIIV}.  Several calculations of $e_{14}$ have been made, and
the results range from\cite{Gironcoli} as small as .02 $C/m^2$
to\cite{Hubner} as large as .22 $C/m^2$.  If we choose one of these
for the value of $e_{14}$ for AlAs and interpolate to obtain $e_{14}$
for ${\mbox{Al}_x\mbox{Ga}_{1-x}\mbox{As}}$ with $x = .3$ (linear
interpolation is thought to be roughly correct\cite{IIIV}) we will
obtain results that range from approximately .11 $C/m^2$ to .18
$C/m^2$.  Since since $e_{14}$ is squared in the final result (Eq.
\ref{eq:msalph}), these uncertainties will be magnified.  Although
this uncertainty results in an overall change in the magnitude of the
coupling, it will not change the functional form of the coupling with
respect to changes in $qd$.  For definiteness, we will choose to work
with a value of $e_{14}$ of .145 $C/m^2$ for
${\mbox{Al}_x\mbox{Ga}_{1-x}\mbox{As}}$ with $x \approx .3$, (which is
close to the value for pure GaAs).  The uncertainty is approximately
.04 $C/m^2$.  When this quantity is squared in Eq. \ref{eq:msalph},
the final result has an uncertainty in scale of about 50\%.

Since the piezoelectric coupling $e_{14}$ is small, it is
clear from the second equation that $\phi$ will be order $e_{14}$
smaller than $u$.  Thus, the first equation will be solved by the
nonpiezoelectric solution discussed above with corrections only at
order $e_{14}^2$.  The mechanical boundary conditions in the
piezoelectric case are\cite{Inge1,Farnell}
\begin{equation}
\label{eq:boundary2}
c_{{\hat{\bf z}}jkl} \partial_l u_k + e_{k{\hat{\bf z}} j} \partial_k \phi = 0.
\end{equation}
Again, this will be satisfied by the nonpiezoelectric solution with
corrections at order $e_{14}^2$.  The electrical boundary condition
that the normal component of electric displacement ${\bf D}$ is
continuous across the surface can be written as\cite{Inge1}
\begin{eqnarray}
  \phi &=& i \frac{v_s^2}{\omega} Z D_{{\hat{\bf z}}} \\ D_{{\hat{\bf z}}} &=&
  e_{{\hat{\bf z}}kl} \partial_l u_k - \epsilon \partial_k \phi
\end{eqnarray}
Where $Z$ is the transverse magnetic wave impedance ($Z = E_x/H_y$) of
the medium above the surface.  The impedance of the adjoining medium
can be written as
\begin{equation}
Z = \frac{i}{v_s \epsilon_0}
\end{equation}
where $\epsilon_0$ is the adjoining medium's dielectric constant.
These conditions can be rewritten as
\begin{equation}
  \label{eq:boundary3}
  \left. 0 = \left(\epsilon_0 q \phi +  
e_{{\hat{\bf z}}kl} \partial_l u_k - \epsilon
  \partial_{{\hat{\bf z}}} \phi \right) \right|_{z=0}
\end{equation}
Here, this boundary condition must be properly treated since it is of
lower order in $e_{14}$.  Thus, we will use the above nonpiezoelectric
solution for $u$ and solve Eqs.  \ref{eq:Poisson} and
\ref{eq:boundary3} for $\phi$.  These two equations in our case can be
recast as
\begin{eqnarray}
  \epsilon \nabla^2 \phi &=& e_{14} (\partial_z u_{xx} + 2 \partial_x
  u_{xz}) \label{eq:Poisson2} \\ 
  0 &=& \left. \left(\epsilon_0 q \phi + e_{14} u_{xx} -
  \epsilon
  \partial_{{\hat{\bf z}}} \phi \right) \right|_{z=0} \label{eq:boundary4}
\end{eqnarray}

The proposed form of solution is
\begin{eqnarray}
  \nonumber \phi &=& \frac{i C e_{14}}{\epsilon} e^{i q ( x - v_s t) }
  \left[ A_1 e^{- \Omega q z -i \varphi} \right. \\ & & ~~~~~~~~~~~~
    \left. + A_2 e^{-\Omega^* q z + i \varphi} + A_3 e^{-q z} \right]
  \label{eq:phiform}
\end{eqnarray}
with $C$ the amplitude of the SAW.  Eq. \ref{eq:Poisson2} immediately
yields the conditions
\begin{equation}
  A_1 = A_2^* =  \frac{\gamma - 2 \Omega}{\Omega^2 - 1}
\end{equation}
Finally, using Eq.\ref{eq:boundary4} yields
\begin{equation}
  A_3 = \frac{- 2}{1+r} \left[\cos \varphi + r \mbox{Re}(A_1 e^{-i
    \varphi}) + \mbox{Re}(\Omega A_1 e^{-i \varphi}) \right]
\end{equation}  
with $r = \epsilon_0/\epsilon \approx \frac{1}{12.5}$.

\begin{figure}[htbp]
  \begin{center}
    \leavevmode
    \epsfxsize=3.0in
    \epsfbox{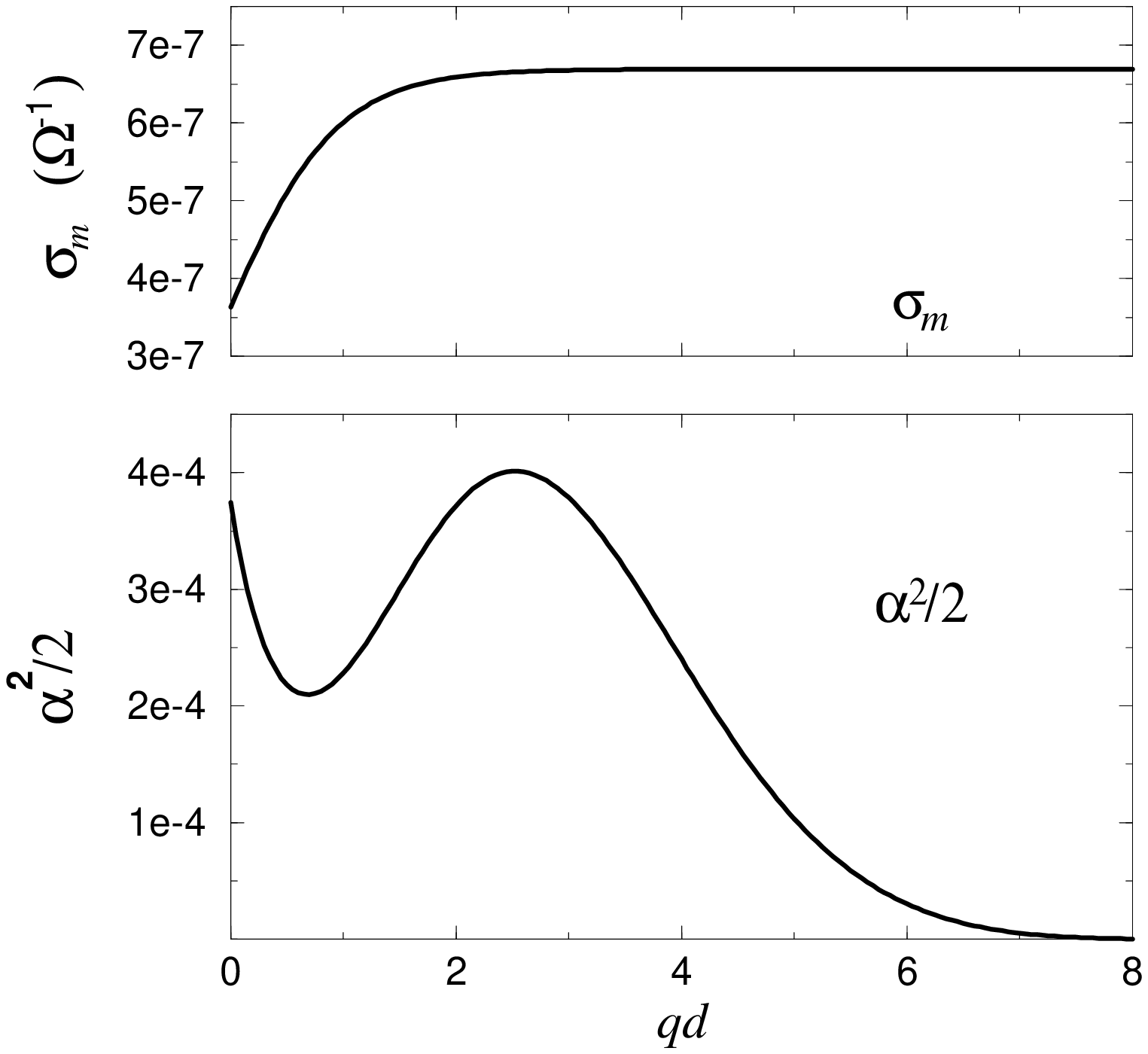}
  \end{center}
  {\centerline{\begin{minipage}[t]{3in} \small{Fig. 2: Coupling
        Constants as a function of $qd$ with $q$ the wavevector and
        $d$ the distance from the 2DEG to the surface.  Top:
        $\sigma_m(qd)$ in units of inverse Ohms.  In this calculation,
        the dielectric constant of the medium
        (${\mbox{Al}_x\mbox{Ga}_{1-x}\mbox{As}}$) is taken to be 12.5.
        Bottom: $\alpha^2(qd)/2$ (dimensionless).  Here, the
        piezoelectric constant is taken as .145 $C/m^2$.  The
        uncertainty in this number results in an uncertainty in the
        scale of approximately $50\%$.  Note that at $qd = 0$ the
        coupling constant is roughly $3.7 \times 10^{-4}$  in good
        agreement with experiment.  }
      \end{minipage}}}
  \label{fig:alpha}
\end{figure}

Once the potential $\phi$ has been determined, this potential is then
treated as the external potential $\phi^{\mbox{\scriptsize{ext}}}$
applied to the 2DEG.  Note that the scalar potential
$\phi^{\mbox{\scriptsize{ind}}}$ then induced by the density
fluctuations in the 2DEG does not change the solution to Eqs
\ref{eq:Poisson} and \ref{eq:boundary3} above since $\nabla^2
\phi^{\mbox{\scriptsize{ind}}} = 0$ everywhere outside of the 2DEG and
$\epsilon_0 q \phi^{\mbox{\scriptsize{ind}}} = \epsilon
  \partial_{{\hat{\bf z}}} \phi^{\mbox{\scriptsize{ind}}} $ 
  at the surface (the solution of such a boundary condition is
  discussed in the appendix).  The form of the potential is given by
  Eq. \ref{eq:aex} where the functional dependence $F$ is given by
\begin{equation}
  F(qd) = 2 |A_1| e^{-\alpha q d} \cos(\beta q z + \phi + \xi) + A_3 e^{-q d}
\end{equation}
where $\Omega = \alpha + \beta i$ and $A_1 = |A_1| e^{-i \xi}$.  Here
we have the values $|A_1| \approx 1.59 $, $\phi + \xi \approx 2.41$,
$A_3 \approx -3.10 $.  Using these values in Eq. \ref{eq:msalph}
yields a the coupling constant $\alpha^2/2$.  The functional
dependence of $\alpha^2/2$ on $qd$ is shown in Fig. \ref{fig:alpha}.
It is clear that the dependence is quite nontrivial.  First of all,
the exponential decay at large $qd$ is roughly proportional to
$e^{-qd}$ rather than $e^{-2qd}$.  This is because, due to the precise
material parameters, the SAW decays into the bulk as $e^{-\alpha q d}$
with $\alpha \approx \frac{1}{2}$.  More importantly, at small $qd$
the coupling seems to oscillate.  The reason for this is roughly that
the boundary condition fixes the strain $u_{xz}$ to be zero at $z =
0$.  Thus $E_{x}(z=0)$ is mainly caused by the surface charge (ie, the
$A_3$ term in Eq. \ref{eq:phiform}).  As $z$ (or $d$) increases, the
effect of the surface charge term quickly decays, but the strain
$u_{xz}$ becomes nonzero so that the coupling decays first, but then
increases.  Finally, at large $qd$, the exponential decay of the SAW
damps out the coupling.

\section{Conclusion and Further Considerations}
\label{sec:conclusion}

This work has focused on surface acoustic waves in AlGaAs coupled to a
2DEG a distance $d$ away from the surface of the sample.  The general
relation (Eq.  \ref{eq:mastersaw}) between the fractional SAW velocity
shift $\Delta v_s/v_s$, the attenuation $\kappa$ and the conductivity
$\sigma_{xx}$ of the 2DEG was derived, and the coefficients
$\alpha^2/2$ and $\sigma_m$ were explicitly calculated as a function
of the product $qd$ of the wavevector $q$ and the distance $d$ to the
surface.

Although Eq. \ref{eq:mastersaw} is very general, the values of the
coefficients $\sigma_m$ and $\alpha^2/2$ are quite dependent on
material parameters.  As discussed in the text and in the appendix,
$\sigma_m$ is dependent only on the velocity of the SAW and on the
effective background dielectric constant in the 2DEG (which is in
general wavevector dependent).  The coupling constant $\alpha^2/2$, on
the other hand, is very sensitive to the details of the sample.  In
this paper we have focused only on a relatively simple model geometry
where the sample is assumed to be a homogeneous slab of AlGaAs to
simplify the solution of the wave problem.  In actual experiments, the
samples are often complicated many layer heterostructures.  In the
relevant experiments\cite{Willettcom}, the bulk of the crystal (below
the 2DEG) is pure GaAs and {\it most} of the crystal between the 2DEG
and the surface is ${\mbox{Al}_x\mbox{Ga}_{1-x}\mbox{As}}$ with $x
\approx .3$.  However, additional thin layers of GaAs are added in
this region, along with Si dopants.  In principle, we could solve the
wave equations for this complicated geometry and apply similar
methods, but in practice such problems can only be solved numerically.
However, since the elastic constants, densities, and dielectric
constants of AlGaAs and GaAs are so similar we suspect that these
heterostructures can be well approximated by the homogeneous system
discussed here.

In using Eq. \ref{eq:mastersaw} to extract $\sigma_{xx}(q,\omega)$
from experimental data, there are several complications.  To begin
with, accurate measurements of the attenuation are extremely
difficult, as are absolute measurements of the
velocity\cite{Willettcom}.  However, measurements of the relative
velocity shift can be made quite accurately.  Another complication is
that the above formula for the velocity shift (Eq.
\ref{eq:mastersaw}) gives the velocity shift $\Delta v_s$ relative to
the velocity of the SAW if the conductivity of the 2DEG were infinite.
In practice, the velocity shift is usually measured relative to the
velocity of the SAW at zero magnetic field.  It is often the case in
high mobility samples (particularly at low frequency) that the
conductivity at zero magnetic field is sufficiently large that it can
be considered infinite and this approximation becomes reasonable.
However, more generally, if the conductivity at zero field is well
known, the resulting measured shift can be appropriately adjusted.

In References
\onlinecite{Weimann,Wixforth,Willett1,Willett2,Willett3,NewWillett},
the parameters $\sigma_m$ and $\alpha^2/2$ are both fit to experiment.
To do this, the dc conductivity is measured and put into Eq.
\ref{eq:mastersaw}, the values of $\sigma_m$ and $\alpha^2/2$ are then
varied until a good fit is obtained to the experimentally measured
values of $\Delta v_s/v_s$ as a function of magnetic field.  There are
several possible problems with this procedure.  To begin with, the
zero frequency (dc) conductivity is expected to be somewhat different
from the finite frequency and wavevector conductivity
$\sigma_{xx}(q,\omega = v_s q)$ that must be used in Eq.
\ref{eq:mastersaw}.  Furthermore, there are indications that due to
large scale inhomogeneities in the sample\cite{Simonnet}, the measured
dc conductivity may not accurately represent the spatial average of
$\sigma_{xx}$.  We thus conclude that these experimental fits of these
parameters to the dc conductivity should be viewed with caution.
Nonetheless the qualitative features of these experiments are
relatively robust and many of the conclusions drawn from these
experiments are relatively independent\cite{Willett2,Willett3} of the
precise value of the fit parameters $\sigma_m$ and $\alpha^2/2$.  A
more careful quantitative analysis of these data is given in Reference
\onlinecite{Simonwave}.

%%%%%%%%%%%%%%%

\vspace*{10pt} \centerline{ACKNOWLEDGMENTS} \vspace*{10pt}

It is a pleasure to acknowledge helpful discussions with Bertrand
Halperin, Bob Willett, Eric Westerberg, and Konstantin Matveev.  The
author is also grateful to G. W. Farnell for sending computer code for
numerically analyzing more complicated SAW problems.  This work is
supported by the National Science Foundation under grant numbers
DMR-94-16910 and DMR-95-23361.

%%%%

\appendix
\section{Coulomb Interaction Near a Dielectric Interface}
\label{app:coulomb}

The Coulomb interaction between electrons in the 2DEG is affected by
the presence of the free surface of AlGaAs since the dielectric
constant of the medium above the surface ($\epsilon_0 \approx 1$) is
much less than the dielectric constant of AlGaAs ($\epsilon \approx
12.5$).  In this appendix, we consider the electrostatic problem of a
charge in a 2DEG a distance $d$ from this AlGaAs/air interface.
Consider a charge $e=1$ placed in the 2DEG at position ${\bf r} = 0$
such that the AlGaAs surface is at the coordinate $z = d$.  It is a
standard result of electrostatics\cite{Jackson} that the electrostatic
potential in the AlGaAs generated by such a charge is given by
\begin{equation}
  \Phi({\bf r}) = \frac{e}{\epsilon}\left(\frac{1}{|{\bf r}|} +
  \left[\frac{\epsilon - \epsilon_0}{\epsilon + \epsilon_0}\right]
    \frac{1}{|{\bf r} + 2 {\hat{\bf z}} d| } \right).
\end{equation}
Here, $|{\bf r} + 2 {\hat{\bf z}} d|$ is the distance from ${\bf r}$ to the
image charge, a distance $d$ away from the surface on the air side.
Restricting ${\bf r}$ to lie in the plane of the 2DEG, and Fourier
transforming, yields
\begin{equation}
  v(k) = \int d^2 r e^{i k \cdot r} \Phi({\bf r})
\end{equation}
which can be evaluated using Eqns. 6.564, 8.411, and 8.469.3 from
Reference \onlinecite{Grad} to yield
\begin{equation}
  v(k) = \frac{2 \pi}{{\epsilon_{\mbox{\tiny{eff}}}} k} 
\end{equation}
where the effective dielectric constant is defined by\cite{Efros}
\begin{equation}
  \frac{1}{{\epsilon_{\mbox{\tiny{eff}}}}} = \frac{1}{\epsilon}\left( 1 +
  \left[\frac{\epsilon- \epsilon_0}{\epsilon + \epsilon_0}\right]
    e^{-2 k d} \right) 
\end{equation}
which can be rewritten in the form of Eq. \ref{eq:eeff}.  Note that
the effective dielectric constant ranges from $\epsilon$ for large $qd
\gg 1$ to $(\epsilon + \epsilon_0)/2$ for $qd \ll 1$.

\end{multicols}

\begin{references}
\bibitem{Grudowski} See also T. W.  Grudowski and M. Gilden, Appl.
  Phys. Lett. {\bf 38}, 412 (1980).  Noted that the experimental
  geometry used in this reference is slightly different from that
  considered here.
\bibitem{endnote2} It should be noted that in Eq. \ref{eq:uform} the
  forms of the solutions assume that the two roots of Eq.
  \ref{eq:Omegaroot} are complex conjugate pairs ($\Omega$ and
  $\Omega^*$).  This is indeed the case for AlGaAs systems.  However,
  for systems with somewhat smaller anisotropy (ie, when $c_{11}$
  closer to $c_{11}'$) the two roots can become real and unequal and
  the form of solution shown here is not correct.
\end{references}
\end{document}